\documentclass[
    ,final            % use final for the camera ready runs
  ]
  {aipproc}

\layoutstyle{6x9}

\usepackage{mathrsfs}
\usepackage{amsfonts}

\def\be{\begin{equation}}
\def\ee{\end{equation}}
\def\bea{\begin{eqnarray}}
\def\eea{\end{eqnarray}}

\begin{document}

\title{Spinflation and Cycling Branes \\ in Warped Throats}

\classification{98.80.Cq, 11.25.-w, 11.25.Uv, 11.25.Mj}
\keywords      {String theory and cosmology, Inflation, D-branes, Flux compactifications}

\author{Damien A. Easson\footnote{Email: \tt damien.easson@durham.ac.uk}}{
  address={Centre for Particle Theory, Department of Mathematical Sciences, Durham University,
Science Laboratories, South Road, Durham, DH1 3LE, UK}
}

\begin{abstract}
The implications of brane motion in angular directions of Calabi-Yau flux compactifications are discussed from the point of view of an observer living on the worldvolume
of the brane and from the point of view of an observer living elsewhere in the three non-compact dimensions. The brane observer
can experience cosmological bounces and cyclic behavior of the scale factor induced by centrifugal angular momentum barriers. Observers 
living elsewhere in the compactification experience marginally prolonged periods of inflation due to large angular momentum (\emph{spinflation}). 
The presence of spinflaton fields (or other fields with non-standard kinetic terms) during inflation may lead to interesting observational signatures in the cosmic microwave background radiation.\end{abstract}

\maketitle

%%%%%%%%%%%%%%%%%%%%%%%%%%%%%%%%%%%%%%%%%%%%
%% MAINMATTER
%%%%%%%%%%%%%%%%%%%%%%%%%%%%%%%%%%%%%%%%%%%%

\section{Introduction}

 The most well studied constructions of brane inflation in string theory are within the context of orientifolds of Calabi-Yau flux compactifications of Type IIB superstring theory.~\footnote{Some early constructions of brane inflation are~\cite{Dvali:1998pa,Alexander:2001ks,Dvali:2001fw,Burgess:2001fx,Brodie:2003qv,Kachru:2003sx}.} 
 In these constructions it is possible to stabilize all phenomenologically dangerous moduli (complex structure and dilaton as well as K{\"a}hler moduli).
 The stabilization involves both fluxes threading cycles of the Calabi-Yau and  non-perturbative elements such as D7-branes (or Euclidean D3-brane instantons) wrapping four-cycles
~\cite{Giddings:2001yu,Kachru:2003aw}.
The fluxes can produce a large warped throat region providing a relatively stable and concrete setting for the study of brane dynamics. 
We consider the motion of a D3-brane probe in this throat both from the point of view of an observer on the brane and from the point of view of an 
observer living \emph{elsewhere} (somewhere in the compactification where the Standard Model can be embedded). In addition to examining radial motion of the brane we
allow the brane to move with angular momentum in the extra dimensions.

 \section{Cycling branes and angular momentum}

  \begin{figure}
  \includegraphics[height=.4\textheight]{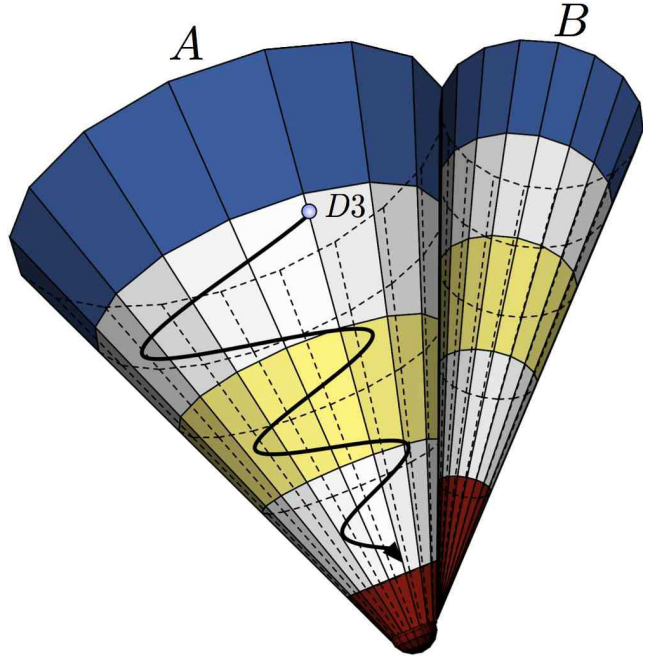}
  \caption{The warped deformed conifold and spinning D3-brane probe. The dual $A$ and $B$ cycles decrease in size from the ultraviolet to the infrared where
  they end in a smooth $S^3$ and zero size $S^2$, respectively. The geometry is composed of $N$ D3 and $M$ D5-branes.}
  \label{KStrc.eps}
\end{figure}

In general, Calabi-Yau spaces are nonsingular; however, for certain values of parameters
they can develop singularities. Locally these singularities appear as a cone, or
``conifold''. When fluxes are introduced this cone can become a warped throat region.
Here we take this throat to be the nonsingular warped deformed conifold (or Klebanov-Strassler (KS) geometry)~\cite{Klebanov:2000hb}.~\footnote{Other possible geometries such as pure $AdS_5$ and the singular warped conifold are considered in~\cite{Easson:2007fz}. } Far from the tip the space is approximated by the conifold $T^{(1,1)} = S^2 \times S^3$ (topologically an $S^2$ fibered over an $S^3$). 
In Calabi-Yau manifolds three-cycles come in Poincar\'e duals. The throat decomposes into two such three-cycles, the $A$ and dual $B$ cycle. 
The geometry is supported by $N$ D3-branes and
 $M$ D5-branes wrapping the $S^2$ of $T^{(1,1)}$ (so-called fractional D3-branes). The background fluxes are composed of both the RR and NS-NS three-forms present in Type IIB superstring theory and
 thread the $A$ and $B$ cycles respectively:
 \be
 \int_A F_3 \sim M \,, \qquad \int_B H_3 \sim -K
 \,,
 \ee
 where $K=N/M$.
 \begin{figure}
  \includegraphics[height=.25\textheight]{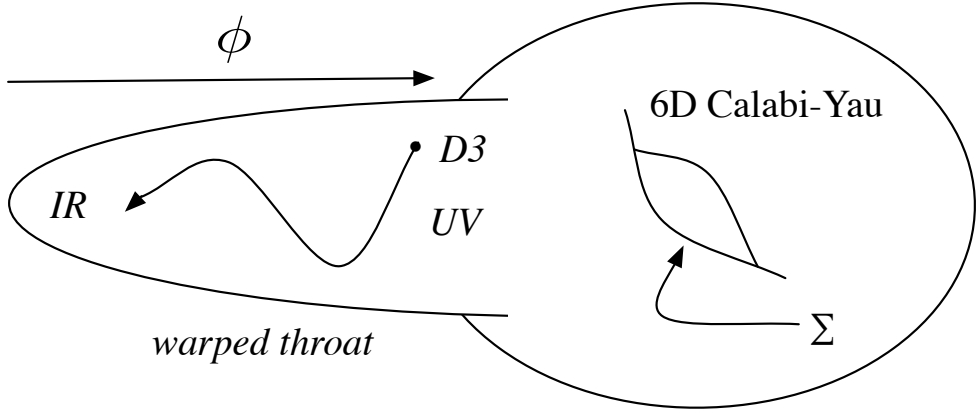}
  \caption{Sketch of Calabi-Yau flux compactification with warped throat region. The position of the D3-brane in the
  throat is parameterized by $\phi$ and $\Sigma$ is a four-cycle. There are wrapped D-branes and orientifold planes not included in this diagram.}
  \label{warped.eps}
\end{figure}
The $A$ cycle ends in a smooth $S^3$ 
(see Figure~\ref{KStrc.eps}). 
The fluxes stabilize the dilaton and complex structure (shape) moduli but do not lift the closed string K\"ahler moduli associated with the volumes of four-cycles in the
compactification~\cite{Giddings:2001yu}.
This stabilization is accomplished by nonperturbative effects such as D7-branes wrapped around the four-cycles $\Sigma_4^{(i)}$ in the space~\cite{Kachru:2003aw}. The KS throat is smoothly glued onto the Calabi-Yau space in the ultraviolet (UV) (see Figure~\ref{warped.eps}).

Consider the dynamics of a D3-brane probe in this background. The D3 brane fills four-dimensional spacetime $\{t, x^1, x^2, x^3 \}$ and appears point-like in the internal space. The metric ansatz is:
 \be
d s_{10}^2\,=\,h^{-1/2}\,g_{\mu \nu} d x^{\mu} d x^{\nu}
\,+\, h^{1/2} \, {\rm g}_{m n} d y^{m} d y^{n}\,,
\ee
where $h$ is the warp factor, $g_{\mu \nu}$ is the metric in 
four-dimensions and $ {\rm g}_{m n}$ is the metric of the internal 
space.

The location of the brane in the extra dimensions is parameterized by a field $\phi$. This field is related to the proper radial throat coordinate $r$ (the proper length with respect to the internal metric (\emph{i.e.}\ ${\rm g}_{rr}=1$)) by $\phi = \sqrt{T_3} r$, where $T_3$ is the brane tension given in terms of the string length, $\alpha' = \ell_s^2$,  by $T_3 = ((2 \pi)^3 \alpha'^2)^{-1}$. In principle the brane is free to move in the five angular coordinates $y^s = \{\psi, \phi_1, \phi_2, \theta_1, \theta_2 \}$. It is the motion in these angular coordinates that is usually ignored and that we focus on here.

The brane motion is described by the Dirac-Born-Infeld (DBI) action plus Wess-Zumino term:
\be
S= -T_3 g_s^{-1}\int{d^4x \, h^{-1}\left[  e^{-3\varphi} \,\sqrt{1-h\,v^2}
  - 1 \right]}\,,
\ee
where $g_s$ is the string coupling, $\varphi$ is the dilaton and $v^2 = g_{mn} \dot y^m \dot y^n$. The warp factor of the KS geometry is:
\be
h(\eta) = (g_s M  \alpha')^2  2^{2/3} \,\zeta^{-8/3}% I(\eta)\,,\\
\int_\eta^\infty{dx\,\frac{x\,\coth{x} -1 }{\sinh^2{x}}\,
\left( \sinh{(2\,x) - 2\,x}\right)^{1/3} }\,.\label{hks}
\ee
Here $\zeta$ is a dimension-full parameter, $[ \zeta^{2/3}] = \ell$, and $\eta$ is the \emph{dimensionless} geometric coordinate parameterizing the throat. 
The geometric variable is related
to the radial coordinate via the transformation:
\be
r = \frac{\zeta^{2/3}}{\sqrt{6}}
\int_0^\eta \frac{\sinh x \, dx}{(\sinh x \, \cosh x - x )^{1/3}}
\,.
\label{ksrdef}
\ee
A tedious calculation yields the surprisingly simple result for  the ten-dimensional Ricci curvature of the KS space:
\be
R = - \frac{4 h'(\eta) +3  h''(\eta) (\coth{\eta} -\eta \, \mbox{csch}^2{\eta})}{\zeta^{4/3} h(\eta )^{3/2} (\cosh{\eta} \sinh{\eta } -\eta )^{1/3}}
\,.
\ee

\begin{figure}
  \includegraphics[height=.3\textheight]{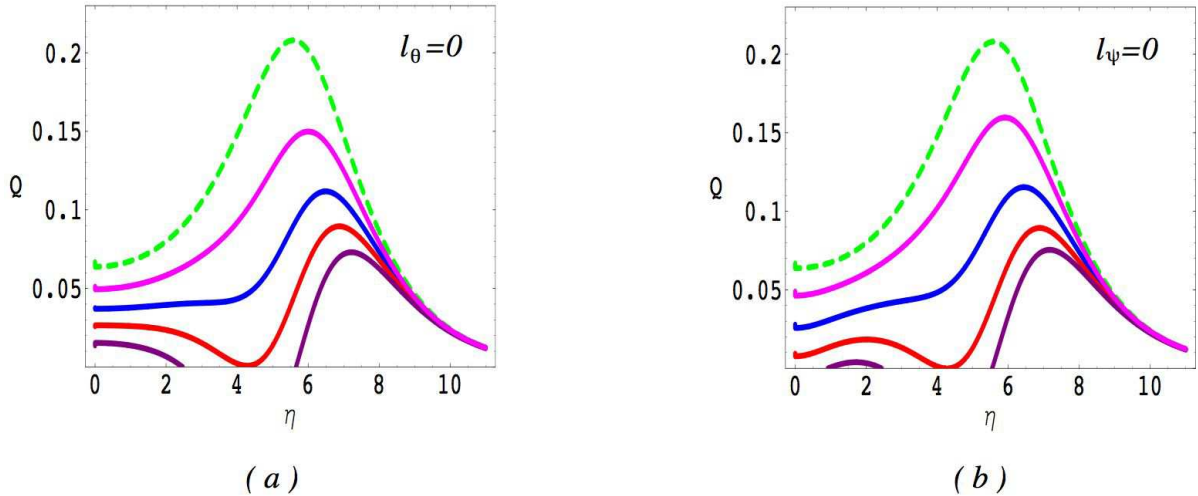}
  \caption{The function $Q$ for a brane moving in the KS background as
the angular momenta is increased from $0$ (top, dashed line) to higher values. 
In $(a)$  we  keep $l_{\theta_1}=0$, and allow $l_\psi$ to
vary, whereas in $(b)$ we keep $l_\psi=0$ and change $l_{\theta_1}$. The zeros of $Q$ correspond to turning points in the brane trajectory.}
  \label{QKSc100.eps}
\end{figure}

As an illustrative example,
we allow the brane to move in the radial $\eta$ direction and in the angular $\psi$ and $\theta_1$ directions:
\be
v^2 = A(\eta)\,\left[ \dot \eta^2   + \dot \psi^2 \right]
+B(\eta) \,\dot\theta_1^2
\,, \ee
where 
\be
A(\eta)= \frac{\zeta^{4/3}}{6K^2}\,, \qquad \quad 
B(\eta) = \frac{\zeta^{4/3} K}{4}\,\left[\cosh^2{(\eta/2)} + 
\sinh^2{(\eta/2)}\right]
\ee 
and
\be
K(\eta) = \frac{\left( \sinh{(2\,\eta) - 2\,\eta}\right)^{1/3}}{2^{1/3}
\,\sinh{\eta}} \,.
\ee
In general, the conserved angular momenta are given by:
\be
\ell_r = \frac{g_{rs} \dot y^2}{\sqrt{1 - hv^2}}
\,,
\ee
corresponding to the conserved q-numbers of the symmetry group.
The conserved angular momenta in our example are
$l_{\psi} = \frac{A\,\dot \psi}{\sqrt{1-h \,v^2}} $, 
$l_{\theta_1} = \frac{B\,\dot \theta_1}{\sqrt{1-h \,v^2}}$,
and the conserved energy (per unit mass) is
\be\label{energyks}
\epsilon= \frac{1}{h} \left[ \sqrt{\frac{1+ h\,\ell^2(\eta)}
{1-h\,g_{\eta\eta}\,\dot\eta^2}} - q \right]
\ee
where 
%\be
$\ell^2(\eta)= 
\frac{l_\psi^2}{A} +\frac{l_{\theta_1}^2}{B} \,.$
%\ee
The time 
evolution of the radial coordinate $\eta$
is
\be\label{etadotKS}
\dot\eta^2 = \frac{\left[  \varepsilon(h\,\varepsilon + 2) 
- \ell^2(\eta)\right]}
{A (h\,\varepsilon+1)^2}= Q
\,.
\ee
The zeros of $Q$ determine turning (bounce) points in the brane trajectory. We plot $Q(\eta)$ for typical values of the parameters in Figure~\ref{QKSc100.eps}.

\subsection{Induced bouncing and cyclic cosmologies}

An observer living on the D3-brane experiences expansion and contraction depending on the motion of the brane in the warped throat. The projected metric in four dimensions may
be written in Friedmann-Robertson-Walker (FRW) form:
\be
ds^2 = h^{-1/2}\left(-(1-h\,v^2)\,dt^2 + dx_i dx^i\right)
= -d\tau^2 + a^2(\tau)dx_i dx^i\,,
\ee
with scale factor 
\be
a(\tau)= h^{-1/4}(\eta(\tau))
\ee
and with brane cosmic time related to the bulk time coordinate 
by
\be
d\tau = h^{-1/4}\sqrt{(1-h\,v^2)}\,dt \,.
\ee
The resulting expansion on the brane is
induced by the warping of the background as in~\cite{Easson:2007fz,Kehagias:1999vr,Kachru:2002kx,sling}. 
From Figure~\ref{QKSc100.eps}, it is clear that the brane observer
can experience bounces in the scale factor $a(\tau)$ either when the brane reaches the end of the smooth KS geometry or when a bounce occurs due
to large angular momentum (zeros of $Q$). When the brane moves down the throat the brane scale factor contracts and when the brane moves up the throat it expands. The expansion and contraction rate is determined by the warp factor in the throat, $h$, and is quantified by the Hubble parameter $H = a^{-1} da/d\tau$.
The lowermost solution in Figure~\ref{QKSc100.eps} (b) is a bound state where the brane bounces back and forth between two zeros of $Q$. This
corresponds to a cyclic cosmology from the brane observer's point of view. 

The above models provide interesting string theory  motivated, time-dependent solutions that yield bounces and
cyclic motion in the scale factor. They therefore provide testing grounds for discovering general physical features more realistic cyclic and bouncing cosmologies
may posses.
There are serious challenges, however, for a cosmological scenario based on the mirage picture to overcome. For example,  because the cosmological expansion is
induced by the background warping, it is unclear how to reheat the Universe or recover Einstein gravity on the brane at late times~\cite{Kachru:2002kx,Linde:2007fr}. Furthermore, 
mirage cosmology branes bouncing through the tip of KS typically have an unacceptably large blue tilt to the scalar spectral index, $n_s$, on super-Hubble scales in the expanding phase~\cite{Brandenberger:2007by}.

\subsection{Spinflation}

Alternative to the mirage picture, one may study cosmological dynamics from the perspective of an observer living elsewhere in the Calabi-Yau (\it e.g. \rm on a stack of $\mathcal{N}$ D3-branes)~\cite{spin}.
In this case the motion of the D3-brane in the throat gives rise to an inflaton field $\phi$ (related to the geometric radial coordinate $\eta$) in the four-dimensional effective theory of the elsewhere 
observer~\cite{Kachru:2003sx}. If the brane is spinning, the conserved angular momentum may appear as 
a field (or several fields, $\phi_m$) in the effective theory obtained by integrating out the internal coordinates:
\bea\label{dbigrav}
S &=& \frac{M_{Pl}^2}{2} \int{d^4x \sqrt{-g} \,R}  \nonumber \\
 && -\,g_s^{-1}\int{d^4x \sqrt{-g}  \left[   f^{-1}\sqrt{ 1
+ f\,g_{mn}g^{\mu\nu}
  \partial_\mu \phi^m\partial_\nu \phi^n} -q\,f^{-1} + V(\phi^m) \right]}\,,
\eea
where the field theory function $f(\phi)$ is related to the warp factor $h(r)$ by $f(\phi) = T_3^{-1} h(\phi/\sqrt{T_3})$.
The four-dimensional Einstein-Hilbert action arises from dimensional reduction of the closed
string sector of the ten-dimensional action~\cite{st}. The fluctuations of the brane are described by the Dirac-Born-Infeld action. 
The Planck mass, $M_{Pl}$ is relate to the internal six-dimensional volume, $V_6$, by 
 $M_{Pl} ^2= V_6/\kappa^2_{10} $, where 
$\kappa_{10}^2 = (2\pi)^7g_s^2\alpha'^4$. The potential $V$ generally arises once the system is coupled to other
 sectors of the theory. In a realistic model the potential can be quite complicated, however, for our purposes it suffices to 
 consider the simple $m^2 \phi^2$ potential.
 
 Assuming an FRW ansatz for the metric and defining 
$\beta = (3\,g_s\,M_{Pl}^2)^{-1}$, the equations of motion can be written
 \begin{eqnarray}
f g_{\phi\phi} \,\dot{\phi}^2 &=& 1-\left(1+\frac{f \ell^2(\phi)}{a^6}\right)
\cdot \left(1+f\left(\frac{ H^2}{\beta}-
 V\right) \right)^{-2}\label{phidot1} \\
\dot{H}&=&-\frac{3 \beta}{2 }\left[ 2   \left(\frac{ H^2}{\beta}-
 	V\right)+f \left(\frac{ H^2}{\beta} - V\right)^2
\right]\cdot  \left(1+f\left(\frac{ H^2}{\beta}
		- V\right) \right)^{-1}\label{Hdot1}
		\,,
\end{eqnarray}
subject to the Friedmann constraint
\be 
H^2 = \beta \left( \frac{1}{f}\left[ \gamma -1 \right] +V \right)
\,,
\ee
where we have defined the Hubble parameter $H \equiv \dot a/a$ and angular momentum $\ell^2(\phi)= g^{rs}\,l_r l_s$, $l_m \equiv a^3\,g_{mn}\,\dot \theta^n\gamma$ and $\gamma$ is the
``lorentz'' factor generated by the non-standard kinetic term for the fields in (\ref{dbigrav}):
\be\label{gammal}
\gamma\,=\,\sqrt{
\frac{
1+f \,\ell^2(\phi)/a^6}{1-f \,g_{\phi\phi} \,\dot{\phi}^2}}\,.
\ee
Equation (\ref{gammal}) places an upper bound on the brane velocity in the radial direction: $|\dot\phi| < \sqrt{f^{-1}}$. The velocity bound
operates independently from the presence of angular momentum and leads to inflation in models with potentials normally too steep to do so~\cite{st}. 
It would appear, however, that it is difficult to build a successful
model of DBI inflation in the KS geometry, as $\gamma$ seemingly grows too large to ignore back reaction effects of the probe brane~\cite{spin}.

From the point of view of the elsewhere observer and in contrast to the brane-bound observer the D3-brane inflaton \it always \rm drives expansion of the
four-dimensional scale factor $a(t)$. The brane probe bounces up and down through the tip of the KS throat 
leading to inflation sourced by the DBI action (\ref{dbigrav}). The bounces are strongly damped by the force due to the potential $V$. If the inflaton couples
to standard model fields (\emph{e.g.} through couplings of the form $g^2 \phi^2 \xi^2$ or $h \phi \bar\psi \psi$) reheating will occur~\cite{Brodie:2003qv}.

The addition of angular momentum alters the form of the acceleration parameter:
\be
\varepsilon\,\equiv\,-\frac{\dot{H}}{H^2} 
\,.
\ee
Using equations  (\ref{phidot1}) and (\ref{Hdot1}) we have:
\be
\varepsilon\,=\,\frac{3\beta}{2 H^2}
\left\{
\left[1+f \left( \frac{3 H^2}{\beta} -V  \right)
\right]\,\dot{\phi}^2+\frac{l^2(\phi)}{a^6}
\left[ 
1+f \left( \frac{3 H^2}{\beta} - V \right)
\right]^{-1} \right\}\label{epsilonpar}
\,.
\ee
Although the angular momentum term adds a positive contribution to $\epsilon$, its dominant effect is to decrease the overall value of $\epsilon$ by decreasing
the brane speed $\dot\phi$. Hence, the addition of angular momentum tends to prolong the inflationary period by allowing the acceleration parameter to remain small
for a slightly longer period of time. The angular momentum is suppressed by $a^{-6}$ and is therefore rapidly diluted during inflation; however,
\emph{spinflaton} fields can help a given inflationary scenario by providing a small number of extra e-foldings of expansion at early times.

It would be interesting to look for observable signatures of spinflaton fields, for example, in the cosmic microwave background radiation. Such a search
requires an understanding of cosmological perturbation theory in multi-field models with non-standard kinetic terms. We have initiated this study 
in \cite{spin}. We found the distinctive feature that non-adiabatic entropy modes and adiabatic curvature perturbations propagate with different speeds.
This leads to a suppression in the conversion of isocurvature to curvature perturbations and
unique observable features in the form of non-gaussianities~\cite{bret}, above and beyond the familiar non-gaussianities associated with DBI inflation~\cite{ast}.

%\section{Discussion}

%%%%%%%%%%%%%%%%%%%%%%%%%%%%%%%%%%%%%%%%%%%%%%%%
%% BACKMATTER
%%%%%%%%%%%%%%%%%%%%%%%%%%%%%%%%%%%%%%%%%%%%%%%%

\begin{theacknowledgments}

It is a pleasure to thank my collaborators R.~Gregory,  D.~Mota, G. Tasinato and  I.~Zavala and the organizers of PASCOS 07.
I am grateful to R.~Brandenberger, C.~Burgess, H.~Firouzjahi, C.~Germani,  L.~Kofman, A.~Linde, L.~McAllister, F.~Quevedo, G.~Shiu, A. Tolley, H.~Tye, B.~Underwood and D.~Wesley for useful discussions.
 This work is supported in part by PPARC and by the EU 6th Framework Marie Curie Research and Training network ``UniverseNet'' (MRTN-CT-2006-035863). DCPT-07/51.
\end{theacknowledgments}


\end{document}